\documentclass[english,conference]{IEEEtran}
\usepackage[T1]{fontenc}
\usepackage[latin9]{luainputenc}
\usepackage{babel}
\usepackage{amsmath}
\usepackage{amsthm}
\usepackage{amssymb}
\usepackage{stmaryrd}
\usepackage[unicode=true,
 bookmarks=true,bookmarksnumbered=true,bookmarksopen=true,bookmarksopenlevel=1,
 breaklinks=false,pdfborder={0 0 1},backref=false,colorlinks=false]
 {hyperref}
\hypersetup{pdftitle={Achieving Super-Resolution in Multi-Rate Sampling Systems via Efficient Semidefinite Programming},
 pdfauthor={Maxime Ferreira Da Costa; Wei Dai},
 pdfsubject={Achieving Super-Resolution in Multi-Rate Sampling Systems via Efficient Semidefinite Programming},
 pdfkeywords={Super-Resolution; Spikes Model; Multirate Sampling Systems; Sparse Polynomials; Semidefinite Programming; Dimensionality Reduction; Exact Frequency Recovery}}

\makeatletter
\theoremstyle{plain}
\newtheorem{thm}{\protect\theoremname}
\theoremstyle{definition}
\newtheorem{defn}[thm]{\protect\definitionname}
\theoremstyle{plain}
\newtheorem{prop}[thm]{\protect\propositionname}
\theoremstyle{plain}
\newtheorem{lem}[thm]{\protect\lemmaname}

\makeatother

\providecommand{\definitionname}{Definition}
\providecommand{\lemmaname}{Lemma}
\providecommand{\propositionname}{Proposition}
\providecommand{\theoremname}{Theorem}

\begin{document}

\title{Achieving Super-Resolution in Multi-Rate Sampling Systems via Efficient
Semidefinite Programming}

\author{\IEEEauthorblockN{Maxime Ferreira Da Costa and Wei Dai}\linebreak{}
\IEEEauthorblockA{Department of Electrical and Electronic Engineering, Imperial College
London, United Kingdom\\Email: \{maxime.ferreira, wei.dai1\}@imperial.ac.uk}}
\maketitle
\begin{abstract}
Super-resolution theory aims to estimate the discrete components lying
in a continuous space that constitute a sparse signal with optimal
precision. This work investigates the potential of recent super-resolution
techniques for spectral estimation in multi-rate sampling systems.
It shows that, under the existence of a common supporting grid, and
under a minimal separation constraint, the frequencies of a spectrally
sparse signal can be exactly jointly recovered from the output of
a semidefinite program (SDP). The algorithmic complexity of this approach
is discussed, and an equivalent SDP of minimal dimension is derived
by extending the Gram parametrization properties of sparse trigonometric
polynomials.
\end{abstract}

\global\long\def\trans{\mathrm{{\scriptscriptstyle T}}}

\global\long\def\herm{\mathrm{{\scriptscriptstyle H}}}

\section{Introduction}

Compressed sensing techniques have proven to be of great interests
for detecting, estimating and denoising sparse signals lying on discrete
spaces. On the practical side, the applications of sparse modeling
are many: single molecule imaging via fluorescence, blind source separation
in speech processing, precise separation of multiple celestial bodies
in astronomy, or super-resolution radaring, are among those. However,
the discrete gridding required by the compressed sensing framework
weaken the recovery performances, and more precisely the \emph{resolution}:
the required minimal separation between two components of the sparse
signal to be efficiently distinguished by an observation process.

In the recent years, a particular enthusiasm has been placed on solving
sparse linear inverse problems over continuous spaces. This paradigm
aims to recover the finite subset of components generating a signal,
and lying in a continuous space, by discrete observations of this
signal, distorted by a kernel function. Considering such approach
raises new concerns, in particular, those problems are commonly infinitely
ill-posed. This primordial issue has been addressed for the spikes
model \cite{Candes2014a,Candes2012,Tang2013} via the mean of total-variation
(or atomic) convex relaxation techniques, reducing the dimensionality
on a dramatic manner. Later on, similar results have been derived
for sparse signals lying on some known subspaces in \cite{Yang2016},
using particular kernel functions \cite{Schiebinger2015}, or via
incoherent multiple measurements in \cite{Yang2014}. Generic performance
in noise have been provided \cite{Bhaskar2013} and specific gradient
search algorithms proposed in \cite{Boyd2015} to efficiently solve
this category of problems.

For the spectral case, a complex time signal $x$ is said to follow
the $s$-spikes model if and only if it reads,
\begin{equation}
x(t)=\sum_{l=1}^{s}\alpha_{l}e^{i2\pi\xi_{l}t},\quad\forall t\in\mathbb{R},\label{eq:SpikeModel}
\end{equation}
where $\xi=\left[\xi_{1},\dots,\xi_{s}\right]^{\trans}\in\mathbb{R}^{s}$
is the vector containing the $s$ spectral components generating the
signal $x$, and $\alpha\in\mathbb{\mathbb{C}}^{s}$ the vector of
their associated complex amplitudes. The frequency estimation problem
is naturally defined as building a consistent estimator $\left(\hat{s},\hat{\xi},\hat{\alpha}\right)$
of the parameters $\left(s,\xi,\alpha\right)$, that are supposed
to be unknown, by $N\in\mathbb{N}$ discrete observations $y\in\mathbb{C}^{N}$
of the time signal $x$.

This problem is obviously ill-posed, and since no assumption is a
priori made on the number of frequencies $\hat{s}$ to estimate, there
are infinitely many triplet $\left(\hat{s},\hat{\xi},\hat{\alpha}\right)$
that are coherent with the observation vector $y$. In particular,
the discrete Fourier transform of $y$ forms a consistent spectral
representation of $x$ by $N$ spectral spikes. Among all those estimators,
the one considered to be optimal, in this context,  will be the one
returning the sparsest spectral distribution, i.e., the one achieving
the smallest $\hat{s}_{0}$. The optimal spectral distribution $\hat{x}_{0}$
can be written as the output of an optimization program taking the
form,\vspace{-0.5cm}

\begin{align}
\hat{x}_{0} & =\arg\min_{\hat{x}\in D_{1}}\left\Vert \hat{x}\right\Vert _{0}\label{eq:RegularL0Problem}\\
\text{subject to} & \phantom{\;=\;}y=\mathcal{F}\left(\hat{x}\right),\nonumber 
\end{align}
where $\hat{x}$ is the spectral distribution of $x$, $\left\Vert \cdot\right\Vert _{0}$
represents the limit of the $p$-norm towards $0$, counting the cardinality
of the support. $D_{1}$ denotes the space of absolutely integrable
spectral distributions, and $\mathcal{F}$ denotes a linear operator
fully determined by the sampling process and linking the spectral
domain to the measurements. Since this program is an NP-hard combinatorial
problem, a common approach consists in relaxing the cardinality cost
function into a minimization of the total-variation norm over the
spectral domain, leading to the convex program,
\begin{align}
\hat{x}_{{\rm TV}} & =\arg\min_{\hat{x}\in D_{1}}\left\Vert \hat{x}\right\Vert _{\mathrm{TV}}\label{eq:RegularL1Problem}\\
\text{subject to} & \phantom{\;=\;}y=\mathcal{F}\left(\hat{x}\right).\nonumber 
\end{align}

The previous works in the literature were mostly studying the regularly
spaced observation model, $y_{k}=x\left(\frac{k}{f}\right)$ for $k\in\left\llbracket 0,N-1\right\rrbracket $.
Under such observations, it has been shown that the relaxation proposed
in (\ref{eq:RegularL1Problem}) is exact in the sense that, under
the minimal separation criterion over the normalized frequencies $\Delta\nu=\min\left\{ \mathrm{frac}\left(\nu_{i}-\nu_{j}\right),i\neq j\right\} \geq\frac{4}{N-1}$
of the sparse spectral distribution $\hat{x}$ to recover, the output
of Programs (\ref{eq:RegularL0Problem}) and (\ref{eq:RegularL1Problem})
are identical. Additionally, Program (\ref{eq:RegularL1Problem})
can be reformulated into a semidefinite program (SDP) of dimension
$N+1$, where coefficients of the optimum define a trigonometric polynomial
$Q_{*}$ locating the frequencies of the original signal over the
unit circle. $Q_{*}$ takes modulus $\left|Q_{*}\left(e^{i2\pi\nu}\right)\right|=1$
whenever $2\pi f\nu=\xi_{l}$ and satisfies $\left|Q_{*}\left(e^{i2\pi\nu}\right)\right|<1$
otherwise. It has been shown in \cite{Tang2013} that this optimality
still holds with high probability when extracting at random a small
number of observations from $y$ and discarding the rest of it. Other
studies revealed that the spectral separation condition can be reduced
\cite{Fernandez-granda2015}, and that this model partially extends
to multidimensional signals \cite{Chi2015,Yang2015}.

In this work, our contribution is focused on extending the previous
results on sparse frequency estimation to the framework of multi-rate
sampling systems (MRSS): the observations $y$ are gathered as the
output of $m$ different uniform samplers, working at different sampling
rates, and potentially desynchronized (the samplers process the time
signal $x$ with some arbitrary delays). According to our knowledge,
this approach is the first to extend super-resolution to such a generic
measurement process. This model is of crucial importance, for instance,
when seeking to achieve joint estimation of sparse signals in distributed
sensor networks. Each node, with limited processing capabilities,
samples at its own rate, a delayed version of a complex signal. Collected
data are then sent and merged at a higher level processing unit, performing
a global estimation of the spectral distribution on a joint manner.
MRSS estimation is also a meaningful step towards a super-resolution
theory from non-uniform sampling. 

In Section \ref{sec:Super-Resolving-MRS}, we show in Proposition
\ref{prop:PolynomialAlignment} that, under certain conditions on
the rates and the delays between the samplers, the ``total-variation''
relaxation of the sparse recovery problem can take a polynomial form
similar to the one described in the original paper \cite{Candes2014a}.
We argue that the model benefits from the same performance guarantees,
and from the optimality. We point out that this direct relaxation
has an arbitrary high complexity, making it unsolvable by standard
convex solvers. In Section \ref{sec:Exact-Dimensionality-Reduction},
a novel exact dimensionality reduction of the semidefinite form (\ref{eq:FullSDP})
is presented in Theorem \ref{thm:BRLSparsePol} by extending the theory
of Gram representation of trigonometric polynomials presented in \cite{Dumitrescu2010}
into the sparse case. We conclude that the dual of the main problem
(\ref{eq:PrimalProblem}) can be reformulated in the compact SDP (\ref{eq:ReducedSDP})
whose dimension is equal to the number of observations.

\section{Super-Resolving Multi-Rate Sampling Systems\label{sec:Super-Resolving-MRS}}

\subsection{Observation model}

An MRSS process on a continuous signal $x$ is parametrized by a set
$\mathbb{A}$ of $m$ distinct grids (or samplers) $\mathcal{A}_{j}$,
$j\in\left\llbracket 1,m\right\rrbracket $. Each grid is identified
with a triplet $\mathcal{A}_{j}=\left(f_{j},\gamma_{j},n_{j}\right)$,
where $f_{j}\in\mathbb{R}^{+}$ is its sampling frequency, $\gamma_{j}\in\mathbb{R}$
its delay (in sample unit), and $n_{j}\in\mathbb{N}$ the number of
measurements acquired by the grid. We assume those intrinsic characteristics
to be known. The output $y_{j}$ of the grid $\mathcal{A}_{j}$ sampling
a signal $x$ following the sparse model described in (\ref{eq:SpikeModel})
reads,
\begin{equation}
y_{j}\left[k\right]=\sum_{l=1}^{s}\alpha_{l}e^{i2\pi\frac{\xi_{l}}{f_{j}}\left(k-\gamma_{j}\right)},\quad k\in\left\llbracket 0,n_{j}-1\right\rrbracket .\label{eq:ObservationModel}
\end{equation}

As explained above, the frequency estimation problem is formulated
as finding the sparsest spectral density jointly matching the observation
vectors $y_{j}$, for all $j\in\left\llbracket 1,m\right\rrbracket $.
This problem takes the same form than the combinatorial minimization
program (\ref{eq:RegularL0Problem}), by specifying the equality constraint
$y=\mathcal{F}\left(\hat{x}\right)$ as follows,
\[
y_{j}=\mathcal{F}_{j}\left(\hat{x}\right),\quad\forall j\in\left\llbracket 1,m\right\rrbracket ,
\]
 where $\mathcal{F}_{j}$ is a linear operator denoting the effect
of the spectral density on the samples uniformly acquired by the grid
$\mathcal{A}_{j}$, and is characterized by,\vspace{-0.45cm}

\begin{align*}
\mathcal{F}_{j}:D_{1} & \to\mathbb{C}^{n_{j}}\\
\hat{x} & \mapsto y_{j}:y_{j}\left[k\right]=\\
 & \phantom{====}\int_{\mathbb{R}}\hat{x}\left(\xi\right)e^{i2\pi\frac{\xi}{f_{j}}\left(k-\gamma_{j}\right)}d\xi,\;\forall k\in\left\llbracket 0,n_{j}-1\right\rrbracket .
\end{align*}

\subsection{Convex relaxation}

We recall that Program (\ref{eq:RegularL0Problem}) is NP-hard in
the general case, due to its combinatorial aspects. The relaxation
described in (\ref{eq:RegularL1Problem}) is introduced and takes
the form,
\begin{align}
\hat{x}_{{\rm TV}} & =\arg\min_{\hat{x}\in D_{1}}\left\Vert \hat{x}\right\Vert _{\mathrm{TV}}\label{eq:PrimalProblem}\\
\text{subject to} & \phantom{\;=\;}y_{j}=\mathcal{F}_{j}\left(\hat{x}\right),\quad\forall j\in\left\llbracket 1,m\right\rrbracket .\nonumber 
\end{align}
Such transform has the advantage to turn the original infinite-combinatorial
problem into a convex problem. However, for practical computation,
convexity often is not enough in order to guarantee a successful resolution
of a program. Indeed, the cost function of (\ref{eq:PrimalProblem})
takes values in $D_{1}$, a space having an uncountable dimension.
Convex optimization theory ensures that this category of programs
can be reformulated into semi-infinite programs \cite{Shapiro}: a
convex optimization program of a finite-dimensional cost function
over an infinite-dimensional set of constraints, using the classic
Lagrangian duality. In our settings, the Lagrange dual problem is,
\begin{align}
c_{*} & =\arg\max\sum_{j=1}^{m}\Re\left(\left\langle y_{j},c_{j}\right\rangle \right)\label{eq:DualProblem}\\
\text{subject to} & \phantom{=\;\;}\left\Vert \sum_{j=1}^{m}\mathcal{F}_{j}^{*}\left(c_{j}\right)\right\Vert _{\infty}\leq1,\nonumber 
\end{align}
where $c=\left[c_{1}^{\trans},\dots,c_{m}^{\trans}\right]^{\trans}$
is the dual variable, and $\mathcal{F}_{j}^{*}$ denotes the adjoint
of the operator $\mathcal{F}_{j}$ for the Euclidean inner products.
Since the original problem is only equally constrained, Slatter's
condition is automatically met, and strong duality holds. This implies
that the optima of the primal problem (\ref{eq:PrimalProblem}) and
its dual (\ref{eq:DualProblem}) are equal. Moreover this equality
appends if and only if $\hat{x}_{\mathrm{TV}}$ is primal optimal,
and $c_{*}$ dual optimal \cite{Boyd2004}.

Letting by $\omega_{j}=\frac{2\pi\xi}{f_{j}}$ the normalized pulsation
of array $\mathcal{A}_{j}$, the expression of the adjoint operator
$\mathcal{F}_{j}^{*}$ allows to reformulate the dual constraint into
a boundedness constraint of a sum of exponential polynomials of the
form,
\begin{align*}
\mathcal{F}_{j}^{*}\left(c_{j}\right) & =\sum_{k=0}^{n_{j}-1}c_{j}\left[k\right]e^{-i\left(k-\gamma_{j}\right)\omega_{j}}\\
 & =e^{i\gamma_{j}\omega_{j}}P_{j}\left(e^{-i\omega_{j}}\right),
\end{align*}
where $P_{j}\in\mathbb{C}^{n_{j}-1}\left[X\right]$ is the dual complex
polynomial related to array $\mathcal{A}_{j}$, and is defined by
$P_{j}\left(z\right)=\sum_{k=0}^{n_{j}-1}c_{j}\left[k\right]z^{k}$.

\subsection{Common grid expansion}

It has been shown in \cite{Candes2014a} that the sparse frequency
recovery problem can take the form of a simple SDP when dealing with
regularly spaced samples. However, those results cannot be transposed
in the MRSS framework, since the dual constrained operator $\sum_{j=1}^{m}\mathcal{F}_{j}^{*}\left(c_{j}\right)$
does not take a polynomial form. As an assumption to bridge this concern,
the sampling process $\mathbb{A}$ is supposed to admit a common supporting
grid, ensuring that the observation samples can be uniformly aligned
at a higher virtual rate. The notion of common supporting grid is
defined bellow. Necessary and sufficient conditions in terms of the
parameters of $\mathbb{A}$ for its existence to hold are stated in
Proposition \ref{prop:ExistenceOfCommonSupportingGrid}. 
\begin{defn}
\label{def:CommonSupportingGrid}A grid $\mathcal{A}_{\#}=\left(f_{\#},\gamma_{\#},n_{\#}\right)$
is said to be a \emph{common supporting grid} for a set of sampling
grids $\mathbb{A}=\left\{ \mathcal{A}_{j}\right\} _{j\in\left\llbracket 1,m\right\rrbracket }$
if and only if the set of samples acquired by the MRSS induced by
$\mathbb{A}$ is a subset of the one acquired by $\mathcal{A}_{\#}$.
In formal terms, the definition is equivalent to $\left\{ \frac{1}{f_{j}}\left(k_{j}-\gamma_{j}\right),\,j\in\left\llbracket 1,m\right\rrbracket ,\,k_{j}\in\left\llbracket 0,n_{j}-1\right\rrbracket \right\} \subseteq\left\{ \frac{1}{f_{\#}}\left(k_{\#}-\gamma_{\#}\right),\,k_{\#}\in\left\llbracket 0,n_{\#}-1\right\rrbracket \right\} $.
The set of common supporting grids of $\mathbb{A}$ is denoted by
$\mathcal{C}\left(\mathbb{A}\right)$. Moreover, a common supporting
grid $\mathcal{A}_{*}=\left(f_{*},\gamma_{*},n_{*}\right)$ for $\mathbb{A}$
is said to be \emph{minimal} if and only it satisfies the minimality
condition, $\forall\mathcal{A}_{\#}\in\mathcal{C}\left(\mathbb{A}\right),\quad n_{*}\leq n_{\#}.$
\end{defn}
\begin{prop}
\label{prop:ExistenceOfCommonSupportingGrid}Given a set of $m$ observation
grids $\mathbb{A}=\left\{ \mathcal{A}_{j}=\left(f_{j},\gamma_{j},n_{j}\right)\right\} _{j\in\left\llbracket 1,m\right\rrbracket }$,
a common supporting grid $\mathcal{A}_{\#}$ exists if and only if
there exist $f_{\#}\in\mathbb{R}^{+}$, $\gamma_{\#}\in\mathbb{R}$,
a set of $m$ positive integers $\left\{ l_{j}\right\} \in\mathbb{N}^{m}$,
and a set of $m$ integers $\left\{ a_{j}\right\} \in\mathbb{\mathbb{Z}}^{m}$
satisfying $f_{\#}=l_{j}f_{j}$ and $\gamma_{\#}=l_{j}\gamma_{j}-a_{j}$
for all $j\in\left\llbracket 1,m\right\rrbracket $. Moreover a common
grid $\mathcal{A}_{*}=\left(f_{*},\gamma_{*},n_{*}\right)$ is minimal,
if and only if, $\gcd\left(\left\{ a_{j}\right\} _{j\in\left\llbracket 1,m\right\rrbracket }\cup\left\{ l_{j}\right\} _{j\in\left\llbracket 1,m\right\rrbracket }\right)=1$,
$\gamma_{*}=\max_{j\in\left\llbracket 1,m\right\rrbracket }\left\{ l_{j}\gamma_{j}\right\} $
and $n_{*}=\max_{j\in\left\llbracket 1,m\right\rrbracket }\left\{ l_{j}\left(n_{j}-1\right)-a_{j}\right\} .$

\vspace{-0.1cm}
\end{prop}
The proof of the above proposition is presented in \cite{FerreiraDaCosta}.
In the following, we assume that $\mathbb{A}$ satisfies the conditions
of Proposition \ref{prop:ExistenceOfCommonSupportingGrid}, and we
denote its minimal common supporting grid by $\mathcal{A}_{*}=\left(f_{*},\gamma_{*},n_{*}\right)$.
The next result shows that, under those circumstances, the dual inequality
constraint in (\ref{eq:DualProblem}) takes a polynomial form.
\begin{prop}
Consider the multi-rate sampling system induced by $\mathbb{A}=\left\{ \mathcal{A}_{j}\right\} _{j\in\left\llbracket 1,m\right\rrbracket }$,
if $\mathcal{C}\left(\mathbb{A}\right)\neq\emptyset$ there exists
a complex polynomial $Q\in\mathbb{C}^{n_{*}-1}\left[X\right]$ such
that $\left\Vert \sum_{j=1}^{m}\mathcal{F}_{j}^{*}(c_{j})\right\Vert _{\infty}=\left\Vert Q(e^{i\omega_{*}})\right\Vert _{\infty}$.\label{prop:PolynomialAlignment}
\end{prop}
\begin{IEEEproof}
The proof of this proposition is direct,
\begin{align*}
\sum_{j=1}^{m}\mathcal{F}_{j}^{*}\left(c_{j}\right) & =\sum_{j=1}^{m}e^{i\gamma_{j}\omega_{j}}P_{j}\left(e^{-i\omega_{j}}\right)\\
 & =e^{i\gamma_{*}\omega_{*}}\sum_{j=1}^{m}e^{ia_{j}\omega_{*}}P_{j}\left(e^{-il_{j}\omega_{*}}\right),
\end{align*}
by replacing $\omega_{j}$ by $l_{j}\omega_{*}$ and $l_{j}\gamma_{j}$
by $\gamma_{*}+a_{j}$ in the second equality, where $\left\{ l_{j}\right\} \in\mathbb{N}^{m}$
and $\left\{ a_{j}\right\} \in\mathbb{\mathbb{Z}}^{m}$ qualify the
minimal common supporting grid $\mathcal{A}_{*}$ of $\mathbb{A}$.
It comes that,
\begin{align*}
\sum_{j=1}^{m}\mathcal{F}_{j}^{*}(c_{j}) & =e^{i\gamma_{*}\omega_{*}}Q\left(e^{-i\omega_{*}}\right),
\end{align*}
where $Q(z)=\sum_{j=1}^{m}z^{-a_{j}}P_{j}\left(z^{bl_{j}}\right)$
is a well defined complex polynomial, since $a_{j}\leq0$ by assumption
on the minimality of $\mathcal{A}_{*}$. Taking the infinite norm
on both sides and noticing its invariance by $\omega_{*}\leftarrow-\omega_{*}$
lead to the desired result.
\end{IEEEproof}
Due to the upscaling effect created by the expansion of $\mathbb{A}$
on a common grid $\mathcal{A}_{*}$, the resulting dual polynomial
$Q$ has a degree $n_{*}-1$ that can be potentially much higher than
the initial degrees of the individual dual polynomials $\left\{ P_{j}\right\} _{j\in\left\llbracket 1,m\right\rrbracket }$.
This fact is illustrated by an example in the end of this section.
However, it is easy to notice that $Q$ is sparse, and that it can
be decomposed into a sum over $N_{*}\leq N=\sum_{j=1}^{m}n_{j}$ monomials.
Let us denote by $q\in\mathbb{C}^{n_{*}}$ the vector containing the
coefficients of $Q(z)=\sum_{k=0}^{n_{*}-1}q_{k}z^{k}$ and call by
$\mathcal{I}\subseteq\left\llbracket 0,n_{*}-1\right\rrbracket $,
the subset of cardinality $N_{*}$ containing the powers of the supporting
monomials. One can write the relation $q=C_{\mathcal{I}}c$, where
$c$ is the dual variable of Problem (\ref{eq:DualProblem}), for
an orthogonal selection matrix $C_{\mathcal{I}}\in\left[0,1\right]^{n_{*}\times N_{*}}$
for the subset $\mathcal{I}$. The matrix $C_{\mathcal{I}}$ can be
directly inferred from the settings of $\mathbb{A}$.

Proposition \ref{prop:PolynomialAlignment} ensures that the dual
constraint of the dual problem described in (\ref{eq:DualProblem})
is equivalent to restrict a complex polynomial to be bounded in modulus
by one around the unit circle $\mathbb{T}$. We recall a result presented
in \cite{Dumitrescu2010} (Corollary 4.25) emerging from the Gram
parametrization theory of complex polynomials which yields,\vspace{-0.4cm}

{\small{}
\begin{equation}
\left\Vert Q(e^{i\omega})\right\Vert _{\infty}\leq1\Leftrightarrow\exists H\,\text{Hermitian s.t. }\begin{cases}
\begin{bmatrix}H & q\\
q^{\herm} & 1
\end{bmatrix}\succeq0\\
\mathcal{T}_{n}^{*}\left(H\right)=e_{1,}
\end{cases}\label{eq:EquivalenceBoundedPolHermitianMatrix}
\end{equation}
}for any $Q\in\mathbb{C}^{n-1}\left[X\right]$, where $\mathcal{T}_{n}^{*}$
is the adjoint to the canonical decomposition of Hermitian Toeplitz
matrices of dimension $n$ $\mathcal{T}_{n}$, and is given by $\mathcal{T}_{n}^{*}\left(H\right)[k]=\mathrm{tr}\left(\Theta_{k}H\right)$,
for $k\in\left\llbracket 0,n\right\rrbracket $, where $\Theta_{k}$
is the elementary Toeplitz matrix equals to $1$ on the $k^{th}$
lower diagonal and zero elsewhere, and where $e_{1}$ is the first
vector of the canonical basis of $\mathbb{C}^{n}$

The semi-algebraic duality (\ref{eq:EquivalenceBoundedPolHermitianMatrix}),
combined with Proposition \ref{prop:PolynomialAlignment}, allows
to rewrite the infinite dimensional constraint of Program (\ref{eq:DualProblem})
into a positivity condition of an Hermitian matrix of dimension $n_{*}+1$
given by,\vspace{-0.33cm}

\begin{align}
c_{*} & =\arg\max\Re\left(\left\langle y,c\right\rangle \right)\label{eq:FullSDP}\\
\text{subject to} & \phantom{\phantom{\;=\;}}\begin{bmatrix}H & C_{\mathcal{I}}c\\
\left(C_{\mathcal{I}}c\right)^{\herm} & 1
\end{bmatrix}\succeq0\nonumber \\
\mathcal{} & \phantom{\;=\;}\mathcal{T}_{n_{*}}^{*}\left(H\right)=e_{1}.\nonumber 
\end{align}
The above problem is nothing but a particular case of the convex relations
studied in \cite{Tang2013}. This ensures that the optimum $q_{*}=C_{\mathcal{I}}c_{*}$
induces a sparse complex polynomial $Q_{*}$ that exactly locates
the frequencies of $x$ by solving $\left|Q_{*}\left(e^{i\omega}\right)\right|=1$
around the unit circle $\omega\in\mathbb{T}$, as long a sufficient
minimal spectral separation discussed in \cite{FerreiraDaCosta} is
respected.

Although semidefinite programs are theoretically solvable and certifiable,
practical attempts to recover the frequencies of the time signal $x$
via Program (\ref{eq:FullSDP}) might fail or return inaccurate results
due to the high dimensionality of the constraints. This is the case
in our settings, the square block matrix in (\ref{eq:FullSDP}) has
a size of $n_{*}+1$, which can be considerably higher than the effective
dimension of the observations $N_{*}$, depending of the settings
of the MRSS defined by $\mathbb{A}$. As for illustration purposes,
suppose a delay-only MRSS, where $\mathbb{A}$ is constituted of $m$
grids given by $\mathcal{A}_{1}=\left(f,0,n\right)$, $\mathcal{A}_{j}=\left(f,-\frac{1}{b_{j}},n\right)$
for all $j\in\left\llbracket 2,m\right\rrbracket $, and where the
$\left\{ b_{j}\right\} _{j\in\left\llbracket 2,m\right\rrbracket }$are
jointly coprime. One has $\mathcal{A}_{*}=\left(\left(\prod b_{j}\right)f,0,\left(\prod b_{j}\right)n\right)$,
leading to a matrix constraint of asymptotic dimension $\Omega\left(b^{m}n\right)$
for some constant $b\in\mathbb{R}^{+}$, while the essential dimension
of the problem remains of order $\mathcal{O}\left(mn\right)$.

\section{Exact Dimensionality Reduction\label{sec:Exact-Dimensionality-Reduction}}

In this section, we show that the original dual problem described
in (\ref{eq:DualProblem}) is equivalent to a similar SDP of size
exactly equal to $N_{*}+1$, which is optimal in those settings. To
this end, we first need to recall some results about Gram parametrization
of trigonometric polynomials.

\subsection{Gram parametrization of trigonometric polynomials}

For every non-zero complex number $z\in\mathbb{C}^{*}$, its $n^{th}$
power vector is defined by $\psi_{n}(z)=\left[1,z,\dots,z^{n}\right]^{\trans}$.
A complex trigonometric polynomial $R\in\bar{\mathbb{C}}^{n}\left[X\right]$
of order $\bar{n}=2n+1$ is a linear combination of complex monomials
with positive and negative exponents absolutely bounded by $n$. Such
polynomial $R$ reads,\vspace{-0.3cm}

\[
R\left(z\right)=\sum_{k=-n}^{n}r_{k}z^{-k},\quad\forall z\in\mathbb{C}^{*}.
\]
Each of such entities can be associated with a Gram set, as defined
in Definition \ref{def:GramSet}. Proposition \ref{prop:GramParametrizationTheorem}
states that this duality holds via a simple linear relation with complex
matrices.
\begin{defn}
A complex matrix $G\in\mathbb{C}^{\left(n+1\right)\times\left(n+1\right)}$
is a \emph{Gram matrix} associated with the trigonometric polynomial
$R$ if and only if,\vspace{-0.1cm}
\[
R\left(z\right)=\psi\left(z^{-1}\right)^{\trans}G\psi\left(z\right),\quad\forall z\in\mathbb{C}^{*}.
\]
Such parametrization is, in general, not unique and we denote by $\mathcal{G}\left(R\right)$
the set of matrices satisfying the above relation. $\mathcal{G}\left(R\right)$
is called \emph{Gram set} of $R$.\label{def:GramSet}
\end{defn}
\begin{prop}
For any complex trigonometric polynomial $R$ of order $\bar{n}=2n+1$,
$G\in\mathcal{G}\left(R\right)$ if and only if the relation,
\[
\mathcal{T}_{\bar{n}}^{*}\left(G\right)=r
\]
holds, where $r\in\mathbb{C}^{\bar{n}}$ is the vector containing
the coefficients of $R$ indexed in $\left\llbracket -n,n\right\rrbracket $.\label{prop:GramParametrizationTheorem}
\end{prop}

\subsection{Compact representation of sparse polynomials}

Up to here, the concept of Gram sets adapts to every complex trigonometric
polynomial. If $R$ is of order $\bar{n}$, it defines a set $\mathcal{G}\left(R\right)$
of matrices in $\mathbb{C}^{n\times n}$. In our context, $R$ has
a sparse monomial support, and Gram representations with compact low-dimensional
structures, reflecting this sparsity, are of crucial interest for
the dimensionality reduction approach. Definition \ref{def:CompactGramRepresentation}
introduces the notion of compact representations.
\begin{defn}
A complex trigonometric polynomial $R$ of order $\bar{n}$ is said
to admit a \emph{compact Gram representation} on a matrix $M\in\mathbb{C}^{n\times m}$,
$m\leq n$, if and only if there exists a matrix $G_{M}\in\mathbb{C}^{m\times m}$
such that the relation,
\begin{align*}
R\left(z\right) & =\psi\left(z^{-1}\right)^{\trans}MG_{M}M^{\trans}\psi\left(z\right)\\
 & =\phi_{M}\left(z^{-1}\right)^{T}G_{M}\phi_{M}\left(z\right),\qquad\forall z\in\mathbb{C}^{*}
\end{align*}
holds, where $\phi_{M}(z)=M^{\trans}\psi(z)$. We denote by $\mathcal{G}_{M}\left(R\right)$
the subset of complex matrices satisfying this property.\label{def:CompactGramRepresentation}
\end{defn}
Although it can be difficult to characterize the set of polynomials
admitting a compact representation on a given matrix $M\in\mathbb{C}^{n\times m}$,
a simple criterion exists for the special case of selection matrices
$C_{\mathcal{I}}$. This criterion is recalled from \cite{Dumitrescu2010}
in Proposition \ref{prop:ExistenceOfCompactRepresentation}.
\begin{prop}
A sparse trigonometric polynomial $R\in\mathbb{\bar{C}}^{n}\left[X\right]$,
supported on $\mathcal{J\subseteq}\left[-n,\dots,n\right]$, admits
a projected representation on a selection matrix $C_{\mathcal{I}}$,
$\mathcal{I}\subseteq\left\llbracket 0,n\right\rrbracket $ if and
only if $\mathcal{J\subseteq\mathcal{I}}-\mathcal{I}$.\label{prop:ExistenceOfCompactRepresentation}
\end{prop}

\subsection{Real bounded lemma for sparse polynomials}

This part aims to demonstrate the novel Theorem \ref{thm:BRLSparsePol},
certifying that, when the polynomial $Q$ is sparse, the condition
$\left\Vert Q\left(e^{i\omega}\right)\right\Vert _{\infty}\leq1$
is equivalent to the existence of a positive Hermitian matrix $S$
(in a similar way as (\ref{eq:EquivalenceBoundedPolHermitianMatrix})),
whose dimension is equal to $N_{*}+1$, the essential dimension of
Problem (\ref{eq:PrimalProblem}). We latter conclude on the existence
of a compact SDP locating the spikes in $\hat{x}$ with exact precision.
The lemma bellow is first required for the demonstration of the main
theorem.
\begin{lem}
\label{lem:JointSparseGramRep}Let $R\in\mathbb{\bar{C}}^{n}\left[X\right]$
and $R'\in\mathbb{\bar{C}}^{n}\left[X\right]$ be two trigonometric
polynomials with common monomial support on $\mathcal{J\subseteq\mathcal{I}}-\mathcal{I}\subseteq\left[-n,\dots,n\right]$.
Let $\mathcal{G}_{\mathcal{I}}\left(R\right)$ and $\mathcal{G}_{\mathcal{I}}\left(R'\right)$
be respectively the Gram compact sets of $R$ and $R'$ on the selection
matrix $C_{\mathcal{I}}$. The inequality $R'\left(e^{i\omega}\right)\leq R\left(e^{i\omega}\right)$
holds for all $\omega\in\mathbb{T}$ if and only if for every two
Hermitian matrices $G\in\mathcal{G}_{\mathcal{I}}\left(R\right)$
and $G'\in\mathcal{G}_{\mathcal{I}}\left(R'\right)$, one has $G'\preceq G$.
\end{lem}
\begin{IEEEproof}
By Proposition \ref{prop:ExistenceOfCompactRepresentation}, the sets
$\mathcal{G}_{\mathcal{I}}\left(R\right)$ and $\mathcal{G}_{\mathcal{I}}\left(R'\right)$
are not empty. Thus, one can find two matrices $G\in\mathcal{G}_{\mathcal{I}}\left(R\right)$
and $G'\in\mathcal{G}_{\mathcal{I}}\left(R'\right)$. The inequality
$R'\left(e^{i\omega}\right)\leq R\left(e^{i\omega}\right)$ holds
for all $\omega\in\mathbb{T}$ if and only if, $0\leq\phi_{C_{\mathcal{I}}}\left(e^{-i\omega}\right)^{\trans}\left(G-G'\right)\phi_{C_{\mathcal{I}}}\left(e^{i\omega}\right)$
for all $\omega\in\mathbb{T}$. Since $\phi_{C_{\mathcal{I}}}\left(e^{-i\omega}\right)=\overline{\phi_{C_{\mathcal{I}}}\left(e^{i\omega}\right)}$
and by noticing that $\left\{ \phi_{C_{\mathcal{I}}}\left(\omega\right),\;\omega\in\mathbb{T}\right\} $
spans the whole space $\mathbb{C}^{N_{*}}$, we conclude that $R'\left(e^{i\omega}\right)\leq R\left(e^{i\omega}\right)$
for all $\omega\in\mathbb{T}$ if and only if $G'\preceq G$.
\end{IEEEproof}
\begin{thm}
\label{thm:BRLSparsePol}Let $P$ and $Q$ be two polynomials from
$\mathbb{C}^{n}\left[X\right]$ with common monomial support on $\mathcal{I}$.
Define the trigonometric polynomial $R\left(z\right)=P\left(z\right)P^{*}\left(z^{-1}\right)$
for all $z\in\mathbb{C}$, and call by $r\in\mathbb{C}^{n+1}$ the
vector of its negative monomial coefficients such that $R$ can be
written under the form $R\left(z\right)=r{}_{0}+\sum_{k=1}^{n}\left(r_{k}z^{-k}+\overline{r_{k}}z^{k}\right)$,
for all $z\in\mathbb{C}^{*}$. Let by $q\in\mathbb{C}^{n+1}$ the
coefficients of $Q$ and define by $u\in\mathbb{C}^{\left|\mathcal{I}\right|}$
the vector satisfying $q=C_{\mathcal{I}}u$. Then the inequality,
\[
\left|Q\left(e^{i\omega}\right)\right|\leq\left|P\left(e^{i\omega}\right)\right|,\quad\forall\omega\in\mathbb{T},
\]
holds if and only if there exists a matrix $S\in\mathbb{\mathbb{\mathbb{C}^{\left|\mathcal{I}\right|\times\left|\mathcal{I}\right|}}}$
satisfying the conditions,
\begin{equation}
\begin{cases}
\begin{bmatrix}S & u\\
u^{H} & 1
\end{bmatrix}\succeq0\\
\mathcal{T}_{n}^{*}\left(C_{\mathcal{I}}SC_{\mathcal{I}}^{\herm}\right)=r.
\end{cases}\label{eq:CompactConstraints}
\end{equation}
\end{thm}
\begin{IEEEproof}
The inequality $\left|Q\left(e^{i\omega}\right)\right|\leq\left|P\left(e^{i\omega}\right)\right|$
is equivalent to $\left|Q\left(e^{i\omega}\right)\right|^{2}\leq\left|P\left(e^{i\omega}\right)\right|^{2}$
for all $\omega\in\mathbb{T}$. Denote by $R$ and $R'$ the two trigonometric
polynomials $R\left(e^{i\omega}\right)=\left|P\left(e^{i\omega}\right)\right|^{2}$
and $R'\left(e^{i\omega}\right)=\left|Q\left(e^{i\omega}\right)\right|^{2}$.
It comes the equivalence with the inequality $R'\left(e^{i\omega}\right)\leq R\left(e^{i\omega}\right)$,
while $R$ and $R'$ are commonly supported by some subset $\mathcal{J}$
satisfying $\mathcal{J}\subseteq\mathcal{I}-\mathcal{I}$.

Let $q\in\mathbb{C}^{n+1}$ be the coefficients of the polynomial
$Q$. Since $R'$ is the square of $Q$, the rank one matrix $qq^{\herm}$
belongs to $\mathcal{G}\left(R'\right)$. Moreover, $q$ is supported
by the subset $\mathcal{I}$, if and only if there exists a $u\in\mathbb{C}^{\left|\mathcal{I}\right|}$
such that $q=C_{\mathcal{I}}u$, and thus if and only if there exists
a matrix $uu^{\herm}\in\mathcal{G}_{\mathcal{I}}\left(R'\right)$.

By application of Lemma \ref{lem:JointSparseGramRep}, an Hermitian
matrix $S\in\mathcal{G}_{\mathcal{I}}\left(R\right)$ satisfying $S\succeq uu^{\herm}$
exists if and only if $R\left(e^{i\omega}\right)\leq R'\left(e^{i\omega}\right)$
for all $\omega\in\mathbb{T}$. We conclude by identification with
a Schur complement that the block matrix inequality in (\ref{eq:CompactConstraints})
holds if and only if $\left|Q\left(e^{i\omega}\right)\right|\leq\left|P\left(e^{i\omega}\right)\right|$,
for all $\omega\in\mathbb{T}$.

In addition, by Proposition \ref{prop:GramParametrizationTheorem},
$S\in\mathcal{G}_{\mathcal{I}}\left(R\right)$ is equivalent to $\mathcal{T}_{n}^{*}\left(C_{\mathcal{I}}SC_{\mathcal{I}}^{\herm}\right)=r$,
which concludes the proof.
\end{IEEEproof}
\medskip{}

Applying Theorem \ref{thm:BRLSparsePol} in the specific case where
$T\left(e^{i\omega}\right)=1$, for all $\omega\in\mathbb{T}$, the
bounded polynomial constraint of Problem (\ref{eq:DualProblem}) verifies
the semidefinite equivalence,{\small{}
\[
\begin{cases}
\left\Vert Q(e^{i\omega})\right\Vert _{\infty}\leq1\\
q=C_{\mathcal{I}}c
\end{cases}\Leftrightarrow\exists S\,\text{Hermitian s.t. }\begin{cases}
\begin{bmatrix}S & c\\
c^{\herm} & 1
\end{bmatrix}\succeq0\\
\mathcal{T}_{n_{*}}^{*}\left(C_{\mathcal{I}}SC_{\mathcal{I}}^{\herm}\right)=e_{1}.
\end{cases}
\]
}where $e_{1}$ is the first vector of the canonical basis of $\mathbb{C}^{n_{*}}$.
Finally, we conclude on our main result, stating that Problem (\ref{eq:DualProblem})
is equivalent to the following reduced SDP,
\begin{align}
c_{*} & =\arg\max\,\Re\left(\left\langle y,c\right\rangle \right)\label{eq:ReducedSDP}\\
\text{subject to} & \phantom{\phantom{\;=\;}}\begin{bmatrix}S & c\\
c^{\herm} & 1
\end{bmatrix}\succeq0\nonumber \\
 & \phantom{\;=\;}\mathcal{T}_{n_{*}}^{*}\left(C_{\mathcal{I}}SC_{\mathcal{I}}^{\herm}\right)=e_{1}.\nonumber 
\end{align}
It is shown in \cite{FerreiraDaCosta} that due to the sparse structure
of $C_{\mathcal{I}}$, the equality constraint in (\ref{eq:ReducedSDP}),
involving vectors in $\mathbb{C}^{n_{*}}$, can be composed in $o\left(N_{*}^{2}\right)$
independent linear forms, involving a total of $\frac{N_{*}\left(N_{*}+1\right)}{2}$
variables, which do not degrade the computational complexity of Program
(\ref{eq:ReducedSDP}). By equivalence, the dual optima $c_{*}$ returned
by Problems (\ref{eq:FullSDP}) and (\ref{eq:ReducedSDP}) are similar.
Consequently, the optimal polynomial $Q_{*}\left(e^{i\omega}\right)$,
locating the spikes in $\hat{x}$, can directly be recovered from
the optimum $c_{*}$ of the compact SDP (\ref{eq:ReducedSDP}) via
the simple linear transform $q_{*}=C_{\mathcal{I}}c_{*}$.

\section{Conclusion}

In this work, we extended the theory of super-resolution from discrete
uniform samples to fit in the more generic framework of multi-rate
sampling systems. We have shown that, under the existence of a virtual
common supporting grid, one can build a dual polynomial locating with
exact precision the frequencies, as long as a minimal separation criterion
is met. The numerical complexity arising from this direct extension
can be arbitrary high. We addressed this issue in the novel Theorem
\ref{thm:BRLSparsePol} by developing an equivalence between Hermitian
matrices and bounded sparse polynomials over the unit circle. We have
derived an equivalent SDP (\ref{eq:ReducedSDP}) of optimal dimension
recovering the signal frequencies.

We reserve for a latter work a deeper exploration of the performances
of this model, including a characterization of the resolution and
spectral range benefits of MRSS, as well as an extension of this theory
to non-uniform sampling systems, by removing the common grid hypothesis,
that we believe to be artificial and unnecessary.\vfill{}

\bibliographystyle{IEEEtran}
\bibliography{../ITW_Bibtex}

\begin{thebibliography}{10}
\providecommand{\url}[1]{#1}
\csname url@samestyle\endcsname
\providecommand{\newblock}{\relax}
\providecommand{\bibinfo}[2]{#2}
\providecommand{\BIBentrySTDinterwordspacing}{\spaceskip=0pt\relax}
\providecommand{\BIBentryALTinterwordstretchfactor}{4}
\providecommand{\BIBentryALTinterwordspacing}{\spaceskip=\fontdimen2\font plus
\BIBentryALTinterwordstretchfactor\fontdimen3\font minus
  \fontdimen4\font\relax}
\providecommand{\BIBforeignlanguage}[2]{{%
\expandafter\ifx\csname l@#1\endcsname\relax
\typeout{** WARNING: IEEEtran.bst: No hyphenation pattern has been}%
\typeout{** loaded for the language `#1'. Using the pattern for}%
\typeout{** the default language instead.}%
\else
\language=\csname l@#1\endcsname
\fi
#2}}
\providecommand{\BIBdecl}{\relax}
\BIBdecl

\bibitem{Candes2014a}
E.~J. Cand{\`{e}}s and C.~Fernandez-Granda, ``{Towards a mathematical theory of
  super-resolution},'' \emph{Communications on Pure and Applied Mathematics},
  vol.~67, no.~6, pp. 906--956, 2014.

\bibitem{Candes2012}
------, ``{Super-resolution from noisy data},'' \emph{Journal of Fourier
  Analysis and Applications}, vol.~19, no.~6, pp. 1229--1254, 2013.

\bibitem{Tang2013}
G.~Tang, B.~N. Bhaskar, P.~Shah, and B.~Recht, ``{Compressed sensing off the
  grid},'' \emph{IEEE Transactions on Information Theory}, vol.~59, no.~11, pp.
  7465--7490, Nov 2013.

\bibitem{Yang2016}
D.~Yang, G.~Tang, and M.~B. Wakin, ``{Super-resolution of complex exponentials
  from modulations with unknown waveforms},'' \emph{arXiv:1601.03712}, 2016.

\bibitem{Schiebinger2015}
G.~Schiebinger, E.~Robeva, and B.~Recht, ``{Superresolution without
  separation},'' \emph{arXiv:1506.03144}, 2015.

\bibitem{Yang2014}
Z.~Yang and L.~Xie, ``{Exact joint sparse frequency recovery via optimization
  methods},'' \emph{arXiv:1405.6585}, 2014.

\bibitem{Bhaskar2013}
B.~N. Bhaskar, G.~Tang, and B.~Recht, ``Atomic norm denoising with applications
  to line spectral estimation,'' \emph{IEEE Transactions on Signal Processing},
  vol.~61, no.~23, pp. 5987--5999, 2013.

\bibitem{Boyd2015}
N.~Boyd, G.~Schiebinger, and B.~Recht, ``{The alternating descent conditional
  gradient method for sparse inverse problems},'' \emph{arXiv:1507.01562},
  2015.

\bibitem{Fernandez-granda2015}
C.~Fernandez-Granda, ``{Super-resolution of point sources via convex
  programming},'' \emph{arXiv:1507.07034}, 2015.

\bibitem{Chi2015}
Y.~Chi and Y.~Chen, ``{Compressive two-dimensional harmonic retrieval via
  atomic norm minimization},'' \emph{IEEE Transactions on Signal Processing},
  vol.~63, no.~4, pp. 1030--1042, 2015.

\bibitem{Yang2015}
Z.~Yang, L.~Xie, and P.~Stoica, ``{Vandermonde decomposition of multilevel
  toeplitz matrices with application to multidimensional super-resolution},''
  \emph{arXiv:1505.02510}, 2015.

\bibitem{Dumitrescu2010}
B.~Dumitrescu, \emph{{Positive Trigonometric Polynomials and Signal Processing
  Applications}}.\hskip 1em plus 0.5em minus 0.4em\relax Springer, 2010.

\bibitem{Shapiro}
A.~Shapiro, ``{Semi-infinite programming, duality, discretization and
  optimality conditions},'' \emph{Optimization}, vol.~58, no.~2, pp. 133--161,
  Feb 2009.

\bibitem{Boyd2004}
S.~P. Boyd and L.~Vandenberghe, \emph{{Convex Optimization}}.\hskip 1em plus
  0.5em minus 0.4em\relax Cambridge University Press, 2004.

\bibitem{FerreiraDaCosta}
M.~{Ferreira Da Costa} and W.~Dai, ``On super-resolution in multirate sampling
  systems,'' under review.

\end{thebibliography}

\end{document}